\documentclass[aps,prd,superscriptaddress,showpacs,preprint]{revtex4}
\usepackage{graphicx, bm}
\usepackage[usenames]{color}

\begin{document}

\draft
\title{$Z'$ resonance and associated $Zh$ production at future\\
 Higgs boson factory: ILC and CLIC
}

\author{A. Guti\'errez-Rodr\'{\i}guez\footnote{alexgu@fisica.uaz.edu.mx}}
\affiliation{\small Facultad de F\'{\i}sica, Universidad Aut\'onoma de Zacatecas\\
             Apartado Postal C-580, 98060 Zacatecas, M\'exico.\\}

\author{ M. A. Hern\'andez-Ru\'{\i}z\footnote{mahernan@uaz.edu.mx}}
\affiliation{\small Unidad Acad\'emica de Ciencias Qu\'{\i}micas, Universidad Aut\'onoma de Zacatecas\\
         Apartado Postal C-585, 98060 Zacatecas, M\'exico.\\}

\date{\today}

\begin{abstract}

We study the prospects of the B-L model with an additional $Z'$ boson to be a Higgs boson factory at
high-energy and high-luminosity linear electron positron colliders, such as the ILC and CLIC, through
the Higgs-strahlung process $e^{+}e^{-}\rightarrow (Z, Z') \to Zh$, including both the resonant and
non-resonant effects. We evaluate the total cross section of $Zh$ and we calculate the total number
of events for integrated luminosities of 500-2000\hspace{0.8mm}$fb^{-1}$ and center of mass energies
between 500 and 3000\hspace{0.8mm}$GeV$. We find that the total number of expected $Zh$ events can reach
$10^6$, which is a very optimistic scenario and it would be possible to perform precision measurements
for both the $Z'$ and Higgs boson in future high-energy $e^+e^-$ colliders experiments.

\end{abstract}


\vspace{5mm}

\maketitle


\section{Introduction}

The discovery of a light scalar boson $H$ of the ATLAS \cite{Aad} and CMS \cite{Chatrchyan} Collaborations at the
Large Hadron Collider (LHC) compatible with a SM Higgs boson \cite{Higgs,Higgs1,Higgs2,Englert,Guralnik} and with
mass around $M_h=125\pm 0.4\mbox{(stat.)}\pm 0.5\mbox{(syst.)}$\hspace{0.8mm}$GeV$ has opened a window to new sectors
in the search for physics beyond the Standard Model (SM). The Higgs boson might be a portal leading to more profound
physics models and even physics principles. Therefore, another Higgs factory besides the LHC, such as the International
Linear Collider (ILC) \cite{Abe,Aarons,Brau,Baer,Asner,Zerwas} and the Compact Linear Collider (CLIC) \cite{Accomando,Dannheim,Abramowicz}
that can study in detail and can precisely determine the properties of the Higgs boson is another important future step
in high-energy and high-luminosity physics exploration.

The existence of a heavy neutral ($Z'$) vector boson is a feature of many extensions
of the Standard Model. In particular, one (or more) additional $U(1)'$ gauge group
provides one of the simplest extensions of the SM. Additional $Z'$ gauge bosons appear
in Grand Unified Theories (GUTs) \cite{Robinett}, Superstring Theories \cite{Green},
Left-Right Symmetric Models (LRSM) \cite{Mohapatra,G.Senjanovic,G.Senjanovic1} and in
other models such as models of composite gauge bosons \cite{Baur}. In particular, it is
possible to study some phenomenological features associated with this extra neutral gauge
boson by considering a $B-L$ (baryon number minus lepton number) model.

The $B-L$ symmetry plays an important role in various physics scenarios beyond the SM:
a) the gauge $U(1)_{B-L}$ symmetry group is contained in a GUT described by a $SO(10)$
group \cite{Buchmuller}. b) the scale of the $B-L$  symmetry breaking is related to
the mass scale of the heavy right-handed Majorana neutrinos mass terms providing the
well-known see-saw mechanism \cite{Mohapatra1} to explain light left-handed neutrino mass.
c) the $B-L$ symmetry and the scale of its breaking are tightly connected to the baryogenesis
mechanism through leptogenesis \cite{Fukugita}.

The $B-L$ model \cite{Basso,Basso0} is attractive due to its relatively simple theoretical
structure, and the crucial test of the model is the detection of the new heavy neutral $(Z')$
gauge boson. The analysis of precision electroweak measurements indicates that the new $Z'$ gauge
boson should be heavier than about 1.2 $TeV$ \cite{Langacker}. On the other hand, recent bounds
from the LHC indicate that the $Z'$ gauge boson should be heavier than about 2 $TeV$ \cite{ATLAS,CMS},
while future LHC runs at 13-14 $TeV$ could increase the $Z'$ mass bounds to higher values, or
may be lucky and find evidence for its presence. Further studies of the $Z'$ properties will
require a new linear collider \cite{Allanach}, which will also allow us to perform precision
studies of the Higgs sector. Detailed discussions on the $B-L$ model can be found in
the literature \cite{Basso,Basso1,Basso2,Basso3,Basso5,Basso6,Lorenzo,Satoshi}.

The Higgs-stralung \cite{Ellis,Ioffe,Lee,Bjorken,Barger} process $e^{+}e^{-} \to Zh$ is one of the main production mechanisms
of the Higgs boson in the future linear $e^+e^-$ colliders experiments, such as the ILC and CLIC. Therefore, after
the discovery of the Higgs boson, detailed experimental and theoretical studies are necessary for checking its
properties and dynamics \cite{Ellis1,Dawson,Klute,Behnke}. It is possible to search for the Higgs boson in the framework of the
B-L model, however the existence of a new gauge boson could also provide new Higgs particle production mechanisms,
which could prove its non-standard origin. In this work, we analyze how the $Z'$ gauge boson of the $U(1)_{B-L}$
model could be used as a factory of Higgs bosons.

Our aim in the present paper is to study the sensitivity of the $Z'$ boson of the B-L model as a Higgs
boson factory through the Higgs-strahlung process $e^{+}e^{-}\rightarrow (Z, Z') \to Zh$, including
both the resonant and non-resonant effects at future high-energy and high-luminosity linear $e^+e^-$
colliders, such as the International Linear Collider (ILC) \cite{Abe} and the Compact Linear Collider
(CLIC) \cite{Accomando}. We evaluate the total cross section of $Zh$ and we calculate the total number of
events for integrated luminosities of 500-2000\hspace{0.8mm}$fb^{-1}$ and center-of-mass energies between
500 and 3000\hspace{0.8mm}$GeV$. We find that the total number of expected $Zh$ events for the $e^+e^-$
colliders is very promising and that it would be possible to perform precision measurements for both the
$Z'$ and Higgs boson in the future high-energy $e^+e^-$ colliders experiments. In addition, we also studied
the dependence of the Higgs signal strengths $(\mu)$ on the parameters $g'_1$ and $\theta_{B-L}$ of the
$U(1)_{B-L}$ model for the Higgs-stralung process $e^+e^- \to Zh$.

This paper is organized as follows. In Section II, we present the theoretical framework.
In Section III, we present the decay widths of the $Z'$ boson in the context of the B-L model.
In Section IV, we present the calculation of the process $e^{+}e^{-}\rightarrow (Z, Z') \to Zh$,
and finally, we present our results and conclusions in Section V.

\vspace{5mm}

\section{Theoretical Framework}

We consider an $SU(2)_L\times U(1)_Y\times U(1)_{B-L}$ model consisting of one doublet $\Phi$ and one
singlet $\chi$ and briefly describe the lagrangian including the scalar, fermion and gauge sector.
The Lagrangian for the gauge sector is given by \cite{Ferroglia,Rizzo,Khalil,Basso6}

\begin{equation}
{\cal L}_g=-\frac{1}{4}B_{\mu\nu}B^{\mu\nu}-\frac{1}{4}W^a_{\mu\nu}W^{a\mu\nu}-\frac{1}{4}Z'_{\mu\nu}Z^{'\mu\nu},
\end{equation}

\noindent where $W^a_{\mu\nu}$, $B_{\mu\nu}$ and $Z'_{\mu\nu}$ are the field strength tensors for $SU(2)_L$, $U(1)_Y$
and $U(1)_{B-L}$, respectively.

The Lagrangian for the scalar sector of the $SU(2)_L\times U(1)_Y\times U(1)_{B-L}$ model is

\begin{equation}
{\cal L}_s=(D^\mu\Phi)^\dagger(D_\mu\Phi) + (D^\mu\chi)^\dagger(D_\mu\chi)-V(\Phi, \chi),
\end{equation}

\noindent where the potential term is \cite{Basso3},

\begin{equation}
V(\Phi, \chi)=m^2(\Phi^\dagger \Phi)+\mu^2|\chi|^2+\lambda_1(\Phi^\dagger \Phi)^2 + \lambda_2|\chi|^4 + \lambda_3(\Phi^\dagger \Phi)|\chi|^2.
\end{equation}

\noindent with $\Phi$ and $\chi$ as the complex scalar Higgs doublet and singlet fields, respectively.
The covariant derivatives for the doublet and singlet are given by \cite{Basso1,Basso2,Basso3}

\begin{eqnarray}
D^\mu\Phi&=&\partial_\mu\Phi + i[gT^aW^a_\mu + g_1YB_\mu + g'_1Y'B'_\mu ]\Phi,\nonumber\\
D^\mu\chi&=&\partial_\mu\chi + i[g_1YB_\mu + g'_1Y'B'_\mu ]\chi,
\end{eqnarray}

\noindent where the doublet and singlet scalars are

\begin{eqnarray}
\Phi= \left(
       \begin{array}{c}
        G^{\pm}\\
      \frac{v+\phi^0 +i G_Z}{\sqrt{2}}
      \end{array}
      \right), \hspace{1cm}
\chi= \left(\frac{v'+\phi^{'0} +i  z'}{\sqrt{2}}\right),
\end{eqnarray}

\noindent with $G^{\pm}$, $G_Z$ and $z'$ the Goldstone bosons of $W^{\pm}$, $Z$ and $Z'$, respectively.

After spontaneous symmetry breaking the two scalar fields can be written as,

\begin{eqnarray}
\Phi= \left(
       \begin{array}{c}
        0\\
      \frac{v+\phi^0}{\sqrt{2}}
      \end{array}
      \right), \hspace{1cm}
\chi= \frac{v'+\phi^{'0}}{\sqrt{2}},
\end{eqnarray}

\noindent with $v$ and $v'$ real and positive. Minimization of Eq. (3) gives

\begin{eqnarray}
m^2+2\lambda_1v^2+\lambda_3vv'^2&=&0,\nonumber\\
\mu^2+4\lambda_2v'^2+\lambda_3v^2v'&=&0.
\end{eqnarray}

To compute the scalar masses, we must expand the potential in Eq. (3) around the minima in
Eq. (6). Using the minimization conditions, we have the following scalar mass matrix:

\begin{eqnarray}
{\cal M}= \left(
       \begin{array}{rr}
        \lambda_1v^2& \frac{\lambda_3vv'}{2}\\
      \frac{\lambda_3vv'}{2}& \lambda_2v'^2
      \end{array}
      \right)
      = \left(
       \begin{array}{rr}
        {\cal M}_{11}& {\cal M}_{12} \\
        {\cal M}_{21}& {\cal M}_{22}
       \end{array}
      \right).
\end{eqnarray}

The expressions for the scalar mass eigenvalues $(m_{H'} > m_h)$ are

\begin{equation}
m^2_{H', h}=\frac{({\cal M}_{11}+{\cal M}_{22})\pm \sqrt{({\cal M}_{11}- {\cal M}_{22})^2+4{\cal M}^2_{12}}}{2},
\end{equation}

\noindent and the mass eigenstates are linear combinations of $\phi^0$ and $\phi^{'0}$, and written as,

\begin{eqnarray}
\left(
       \begin{array}{c}
        h\\
        H'
      \end{array}
      \right)
= \left(
       \begin{array}{rr}
       \cos\alpha& -\sin\alpha\\
       \sin\alpha& \cos\alpha
      \end{array}
      \right)
\left(
       \begin{array}{c}
        \phi^0\\
        \phi^{'0}
      \end{array}
      \right),
\end{eqnarray}

\noindent where $h$ is the SM-like Higgs boson. The scalar mixing angle, $\alpha$ can be expressed as

\begin{equation}
\tan{(2\alpha)}=\frac{2{\cal M}_{12}}{{\cal M}_{11}-{\cal M}_{22}}=\frac{\lambda_3vv'}{\lambda_1v^2-\lambda_2v'^2}.
\end{equation}

In the Lagrangian of the $SU(2)_L\times U(1)_Y\times U(1)_{B-L}$ model, the terms for the interactions between
neutral gauge bosons $Z, Z'$ and a pair of fermions of the SM can be written in the form \cite{Basso6,Lorenzo}

\begin{equation}
{\cal L}_{NC}=\frac{-ig}{\cos\theta_W}\sum_f\bar f\gamma^\mu\frac{1}{2}(g^f_V- g^f_A\gamma^5)f Z_\mu + \frac{-ig}{\cos\theta_W}\sum_f\bar f\gamma^\mu\frac{1}{2}(g^{'f}_V- g^{'f}_A\gamma^5)f Z'_\mu.
\end{equation}

\noindent From this Lagrangian we determine the expressions for the new couplings of the $Z, Z'$ bosons with the SM fermions, which are given by

\begin{eqnarray}
g^f_V&=&T^f_3\cos\theta_{B-L}-2Q_f\sin^2\theta_W\cos\theta_{B-L}+\frac{2g'_1}{g}\cos\theta_W \sin\theta_{B-L},\nonumber\\
g^f_A&=&T^f_3\cos\theta_{B-L},\\
g^{'f}_V&=&-T^f_3\sin\theta_{B-L}-2Q_f \sin^2\theta_W \sin\theta_{B-L}+\frac{2g'_1}{g}\cos\theta_W \cos\theta_{B-L},\nonumber\\
g^{'f}_A&=&-T^f_3\sin\theta_{B-L},
\end{eqnarray}

\noindent where $g=e/\sin\theta_W$ and $\theta_{B-L}$ is the $Z-Z'$ mixing angle. The current bound on this parameter is $|\theta_{B-L}|\leq 10^{-3}$ \cite{Data2014}. In the decoupling limit, that is to say, when   $g'_1=0$ and $\theta_{B-L}=0$ the couplings of the SM are recovered.

\section{The decay widths of $Z'$ in the B-L model}

In this section we present the new decay widths of the $Z'$ boson \cite{Leike,Langacker,Robinet,Barger1}
in the context of the B-L model which we need in the calculation of the cross section for the process
$e^+e^- \to Zh$. The $Z'$ partial decay widths involving vector bosons and the scalar boson are

\begin{equation}
\Gamma(Z' \to W^+W^-)=\frac{G_F M^2_W}{24\pi\sqrt{2}} \cos^2\theta_W\sin^2\theta_{B-L}M_{Z'}\biggl(\frac{M_{Z'}}{M_Z}\biggr)^4
\biggl(1-4\frac{M^2_W}{M^2_{Z'}}\biggr)^{1/3}
\biggl[1+20\frac{M^2_W}{M^2_{Z'}}+12\frac{M^4_W}{M^4_{Z'}}\biggr],
\end{equation}

\begin{equation}
\Gamma(Z' \to Zh)=\frac{G_F M^2_ZM_{Z'}}{24\pi\sqrt{2}}\sqrt{\lambda} \biggl[\lambda+12\frac{M^2_Z}{M^2_{Z'}}\biggr]
\biggl[f(\theta_{B-L})\cos\alpha - g(\theta_{B-L})\sin\alpha \biggr]^2,
\end{equation}

\noindent where

\begin{eqnarray}
\lambda\biggl(1, \frac{M^2_Z}{M^2_{Z'}}, \frac{M^2_h}{M^2_{Z'}}\biggr)&=&1+\biggl(\frac{M^2_Z}{M^2_{Z'}}\biggr)^2+\biggl(\frac{M^2_h}{M^2_{Z'}}\biggr)^2-2\biggl(\frac{M^2_Z}{M^2_{Z'}}\biggr)
-2\biggl(\frac{M^2_h}{M^2_{Z'}}\biggr)-2\biggl(\frac{M^2_Z}{M^2_{Z'}}\biggr)\biggl(\frac{M^2_h}{M^2_{Z'}}\biggr),\nonumber\\
f(\theta_{B-L})&=&\biggl(\frac{4M^2_Z}{v^2}-g'^2_1\biggr)\sin(2\theta_{B-L})+\biggl(\frac{4g'_1M_Z}{v}\biggr)\cos(2\theta_{B-L}),\\
g(\theta_{B-L})&=&4g'^2_1\biggl(\frac{v'}{v}\biggr)\sin(2\theta_{B-L}).\nonumber
\end{eqnarray}

\noindent The vacuum expectation value $v'$ is taken as  $v'=2$\hspace{0.8mm}$TeV$, while $\alpha=\frac{\pi}{9}$
for the Higgs mixing parameter in correspondence with Refs. \cite{Aad,Chatrchyan,Basso4,Khalil}. In our analysis
we take $v = 246\hspace{0.5mm}GeV$ and constrain the other scale, $v'$, by the lower bounds imposed on the mass
of the extra neutral gauge boson $Z'$. The mass of the $Z'$ and of the heavy neutrinos depend on $v'$, and should
be related to it, while the Higgs masses depend on the angle $\alpha$, the value of which is completely
arbitrary.

\noindent Finally, the decay width of the $Z'$ boson to fermions is given by

\begin{equation}
\Gamma(Z' \to f\bar f)=\frac{2G_F}{3\pi\sqrt{2}}N_fM^2_ZM_{Z'}\sqrt{1-4\biggl(\frac{M^2_f}{M^2_{Z'}}\biggr)}\Biggl[(g'^f_V)^2     \biggl\{1+2\biggl(\frac{M^2_f}{M^2_{Z'}}\biggr)\biggr\}+(g'^f_A)^2 \biggl\{1-4\biggl(\frac{M^2_f}{M^2_{Z'}}\biggr)\biggr\}\Biggr],
\end{equation}

\noindent where $N_f$ is the color factor ($N_f=1$ for leptons, $N_f=3$ for quarks) and the couplings
$g'^f_V$ and $g'^f_A$ of the $Z'$ boson with the SM fermions are given in Eq. (14).

\section{The total cross section of $e^+e^- \to Zh$ in the B-L model}

In this section, we calculate the Higgs production cross section via the process $e^+e^- \to Zh$
in the context of the B-L model at future high-energy and high luminosity linear electron-positron
colliders, such as the ILC and CLIC.

The Feynman diagrams contributing to the process $e^+e^- \to (Z, Z') \to Zh$ are shown in Fig. 1. The
expressions for the total cross section of the Higgs-strahlung process for the different contributions,
that is to say SM, B-L and SM-(B-L), respectively, can be written in the following compact form:\\

\begin{eqnarray}
\sigma(e^+e^- \to Zh)_{tot}&=&\frac{G^2_F M^4_Z}{24\pi}\biggl[(g^e_V)^2+ (g^e_A)^2\biggr]\frac{s \sqrt{\lambda}[\lambda +12M^2_Z/s]}{[(s-M^2_Z)^2+M^2_Z\Gamma^2_Z]}\nonumber\\
&+&\frac{G^2_F M^6_Z}{384\pi}\biggl[(g^{'e}_V)^2+ (g^{'e}_A)^2\biggr]\frac{s \sqrt{\lambda}[\lambda +12M^2_{Z'}/s]}{M^2_{Z'}[(s-M^2_{Z'})^2+M^2_{Z'}\Gamma^2_{Z'}]}\nonumber\\
&\times&\biggl[f(\theta_{B-L})\cos\alpha - g(\theta_{B-L})\sin\alpha\biggr]^2\\
&+&\frac{G^2_F M^6_Z}{12\pi}\biggl[g^e_V g^{'e}_V+ g^e_A g^{'e}_A\biggr] s \sqrt{\lambda}
\biggl [\frac{1}{M^2_Z}(\lambda + 12M^2_Z/s) \nonumber\\
&+& \frac{1}{M^2_{Z'}}(\lambda + 6(M^2_Z-M^2_{Z'})/s)+ \frac{s\lambda}{8M^2_Z M^2_{Z'} }(\lambda - 12M^2_Z/s)\biggr ]\nonumber\\
&\times&\frac{[(s-M^2_Z)(s-M^2_{Z'}) + M_Z M_{Z'}\Gamma_Z\Gamma_{Z'}]}{[(s-M^2_Z)^2+M^2_Z\Gamma^2_Z][(s-M^2_{Z'})^2+M^2_{Z'}\Gamma^2_{Z'}]}\nonumber\\
&\times&\biggl[f(\theta_{B-L})\cos\alpha - g(\theta_{B-L})\sin\alpha\biggr],\nonumber
\end{eqnarray}

\noindent where

\begin{equation}
\lambda\biggl(1, \frac{M^2_Z}{s}, \frac{M^2_h}{s}\biggr)=\biggl(1-\frac{M^2_Z}{s}-\frac{M^2_h}{s}\biggr)^2-4\frac{M^2_ZM^2_h}{s^2},\nonumber\\
\end{equation}

\noindent is the usual two-particle phase space function, while $g^e_V$, $g^e_A$, $g'^e_V$, $g'^e_V$,
$f(\theta_{B-L})$ and $g(\theta_{B-L})$ are given in Eqs. (13), (14) and (17), respectively.

The expression given in the first term of Eq. (19) corresponds to the cross section with the exchange of the $Z$ boson,
while the second and third term come from the contributions of the B-L model and of the interference, respectively.
The SM expression for the cross section of the reaction $e^+e^- \to Zh$ can be obtained in the decoupling
limit, that is to say, when $\theta_{B-L}= 0$ and $g'_1=0$, in this case the terms that depend on $\theta_{B-L}$
and $g'_1$ in (19) are zero and (19) is reduced to the expression given in Refs. \cite{Ellis,Barger} for the
standard model.

\section{Results and Conclusions}

\subsection{$Z'$ resonance and associated $Zh$ production in the B-L model}

In this section we evaluate the total cross section of the Higgs-strahlung process $e^+e^- \to (Z, Z') \to Zh$ in the context of the B-L model at next generation linear $e^+e^-$ colliders such as the ILC and CLIC. Using the following
values for numerical computation \cite{Data2014}: $\sin^2\theta_W=0.23126\pm 0.00022$, $m_\tau=1776.82\pm 0.16\hspace{0.8mm}MeV$, $m_b=4.6\pm 0.18\hspace{0.8mm}GeV$, $m_t=172\pm 0.9\hspace{0.8mm}GeV$, $M_W=80.389\pm 0.023\hspace{0.8mm}GeV$, $M_Z=91.1876\pm 0.0021\hspace{0.8mm}GeV$, $\Gamma_Z=2.4952\pm 0.0023\hspace{0.8mm}GeV$, $M_h=125\pm 0.4$ and considering the most recent limit from LEP \cite{Cacciapaglia}:

\begin{equation}
\frac{M_{Z'}}{g'_1}\geq 7\hspace{0.8mm}TeV,
\end{equation}

\noindent in our numerical analysis, we obtain the total cross section $\sigma_{tot}=\sigma_{tot}(\sqrt{s}, M_{Z'}, g'_1)$.
Thus, in our numerical computation, we will assume $\sqrt{s}$, $M_{Z'}$ and $g'_1$ as free parameters.

We do not consider the process $e^+e^- \to (Z, Z') \to ZH'$ \cite{Basso5} in our study since in major parts of the $U(1)_{B-L}$
model parameter space the higgs boson $H'$ is quite heavy, and it is difficult to detected the process $e^+e^- \to ZH'$
when the relevant mechanism is $e^+e^- \to Zh$.

In Figs. 2-3 we present the total decay width of the $Z'$ boson as a function of $M_{Z'}$ and the new $U(1)_{B-L}$
gauge coupling $g'_1$, respectively, with the other parameters held fixed to three different values. From Fig. 2,
we see that the  total width of the $Z'$ new gauge boson varies from a few to hundreds of $GeV$ over a mass range
of $500\hspace{0.8mm}GeV \leq M_{Z'} \leq 3000\hspace{0.8mm}GeV$, depending on the value of $g'_1$. In the case of Fig. 3,
a similar behavior is obtained in the range $0 \leq g'_1 \leq 1$ and depends on the value of $M_{Z'}$. The branching
ratios versus $Z'$ mass are given in Fig. 4 for different channels, that is to say, $BR(Z' \to f\bar f)$, $BR(Z' \to Zh)$
and $BR(Z' \to W^+W^-)$, respectively. In this figure the $BR(Z' \to f\bar f)$ is the sum of all BRs for the decays into
fermions. We consider $\theta_{B-L}=10^{-3}$, $g'_1=0.5$ and $500\hspace{0.8mm}GeV \leq M_{Z'} \leq 3000\hspace{0.8mm}GeV$.

To illustrate our results on the sensitivity of the $Z'$ gauge boson of the B-L model as a Higgs boson factory
through the Higgs-strahlung process $e^+e^- \to (Z,Z') \to Zh$, including both the resonant and non-resonant
effects at future high-energy and high luminosity linear $e^+e^-$ colliders, such as the International Linear
Collider (ILC) and the Compact Linear Collider (CLIC), we present the total cross section in Figs. 5-11.

In Fig. 5, we show the cross section $\sigma(e^+e^- \to Zh)$ for the different contributions as a function of
the center-of-mass energy $\sqrt{s}$ for $\theta_{B-L}=10^{-3}$ and $g'_1=0.5$: the solid line correspond to the
first term of Eq. (19), where in the $U(1)_{B-L}$ model the couplings $g^f_V$ and $g^f_A$ of the SM gauge boson
$Z$ to electrons receive contributions of the $U(1)_{B-L}$ model. The dashed line corresponds to the second term
of Eq. (19), that is to say, is the pure B-L contribution. Finally, the dot-dashed line corresponds to the total
cross section of the process $\sigma(e^+e^- \to Zh)$. From Figure 5, we can see that the cross section corresponding
to the first term of the Eq. (19) decreases for large $\sqrt{s}$, whereas in the case of the cross section of the B-L model and the total cross section, respectively, these are increased for large values of the center-of-mass energy, reaching its maximum value at the resonance $Z'$ boson, that is to say, $\sqrt{s}=1500$\hspace{0.8mm}$GeV$.

To see the effects of $g'_1$, the free parameter of the B-L model on the process $e^+e^- \to (Z,Z') \to Zh$,
we plot the relative correction $\delta\sigma/\sigma_{SM}=(\sigma_{tot}-\sigma_{SM})/\sigma_{SM}$ as a function
of $g'_1$ for $M_{Z'}=1500, 2000, 2500\hspace{0.8mm}GeV$ and $\sqrt{s}=1500, 2000, 2500\hspace{0.8mm}GeV$ in
Fig. 6. We can see that the relative correction reaches its maximum value between $0.1\leq g'_1\leq 2.5$ and
remains almost constant as $g'_1$ increases.

The deviation of the cross section in our model from the SM one, $\delta\sigma/\sigma_{SM}$ is depicted in Fig. 7
as a function of $M_{Z'}$ for $\sqrt{s}=1500\hspace{0.8mm}GeV$ and three values of the $g'_1$, new gauge coupling.
Figure 7 shows that the relative correction is very sensitive to the gauge  boson mass $M_{Z'}$ and for the gauge
parameter $g'_1= 0.2, 0.5, 0.8$, the peak of the total cross section emerges when the heavy gauge boson mass
approximately equals $M_{Z'}=1500, 1450, 1300\hspace{0.8mm}GeV$, respectively. Thus, in a sizeable parameter region
of the B-L model, the new heavy gauge boson $Z'$ can produce a significant signal, which can be detected in future
ILC and CLIC experiments.

We plot the total cross section of the reaction $e^+e^- \to Zh$ in Figure 8 as a function of the center of
mass energy, $\sqrt{s}$ for the values of the heavy gauge boson mass of $M_{Z'}= 1500, 2000, 2500$\hspace{1mm}$GeV$
and $\theta_{B-L}=10^{-3}$, $g'_1=0.2$, respectively. In this figure we observed that for $\sqrt{s}=M_{Z'}$,
the resonant effect dominates the Higgs particle production. A similar analysis was performed in Fig. 9, but
in this case $\theta_{B-L}=10^{-3}$ and $g'_1=0.5$. In both figures we show that the cross section is sensitive
to the free parameters. Comparing Figs. 8 and 9, we observe that the height of the resonances is the same in both
Figures, but the resonances are broader for larger $g'_1$ values, as the total width of the $Z'$ boson increases
with $g'_1$, as it is shown in Fig.2.

Finally, in Fig. 10 we use the currents values of $M_{Z'}$ and $\theta_{B-L}$, as well as the value
of the coupling constant $g'_1$ and center of mass energy $\sqrt{s}$ of the collider to obtain contour plot
3D for the total cross section $\sigma_{tot}=\sigma_{tot}(\sqrt{s}, M_{Z'}, g'_1)$ of the process $e^+e^- \to Zh$
for $M_h=125$\hspace{0.8mm}$GeV$ and $\theta_{B-L}=10^{-3}$. In this figure the resonance peaks for the boson
$Z'$ are evident for the entire range of allowed parameters of the $U(1)_{B-L}$ model.

\begin{table}[!ht]
\caption{Total production of ZH in the B-L model for $M_{Z'}= 1500, 2000, 2500$\hspace{0.8mm}$GeV$, ${\cal L} = 500, 1000, 2000$\hspace{0.8mm}$fb^{-1}$
(1st, 2nd, 3rd number, respectively, in the last 3 columns), $M_H=125$\hspace{0.8mm}$GeV$, $g'_1=0.5$ and $\theta_{B-L}=10^{-3}$.}
\begin{center}
 \begin{tabular}{cccc}
\hline\hline
\multicolumn{4}{c}{${\cal L}=500, 1000, 2000\hspace{0.8mm}fb^{-1}$}\\
 \hline
 \cline{1-4} $\sqrt{s}$     & $M_{Z'}=1500$\hspace{0.8mm}$GeV$          & $M_{Z'}=2000$\hspace{0.8mm}$GeV$  & $M_{Z'}=2500$\hspace{0.8mm}$GeV$\\
 \hline
500                         &  85 131; 170 263; 340 526                                    &  44 609; 89 219; 178 439               &   34 747; 69 493; 138 987 \\
1000                        &  155 482; 310 964; 621 928                                   &  33 523; 67 047; 134 094               &   15 339; 30 678; 61 355  \\
1500                        &  1 234 000; 2 460 000; 4 930 000                             &  75 192; 150 384; 300 768              &   18 004; 36 008; 72 016   \\
2000                        &  92 640; 185 282; 370 564                                    &  396 490; 792 980; 1 580 000           &   42 224; 84 449; 168 899  \\
2500                        &  20 276; 41 534; 83 069                                      &  52 144; 104 288; 208 577              &   163 538; 327 076; 654 151 \\
3000                        &  8 243; 16 487; 32 974                                       &  12 721; 25 442; 50 885                &   32 173; 64 346; 128 693  \\
\hline\hline
\end{tabular}
\end{center}
\end{table}

From Figs. 5-10, it is clear that the total cross section is sensitive to the value of the gauge boson mass $M_{Z'}$,
center-of-mass energy $\sqrt{s}$ and $g'_1$, the new $U(1)_{B-L}$ gauge coupling, increases with the collider energy
and reaching a maximum at the resonance of the $Z'$ gauge boson. As an indicator of the order of magnitude, we present
the $Zh$ number of events in Table I, for several gauge boson masses, center-of-mass energies and $g'_1$ values and for
a luminosity of ${\cal L}=500, 1000, 2000\hspace{0.8mm}fb^{-1}$. We find that the possibility of observing the process
$e^+e^- \to (Z, Z') \to Zh$ is very promising as shown in Table I, and it would be possible to perform precision
measurements for both the $Z'$ and Higgs boson in the future high-energy linear $e^+e^-$ colliders experiments. We
observed in Table I that the cross section rises once the threshold for $Zh$ production is reached, with the energy,
until the $Z'$ is produced resonantly at $\sqrt{s}=1500, 2000$ and 2500\hspace{1mm}$GeV$, respectively, for the three
cases. Afterwards it decreases with rising energy due to the $Z$ and $Z'$ propagators. Another promising production mode
for studying the $Z'$ boson and Higgs boson properties is $e^+e^- \to (\gamma, Z, Z') \to t\bar th$ \cite{Gutierrez}.

\subsection{The Higgs signal strengths in the B-L model}

Considering the Higgs boson decay channels, the Higgs signal strengths can be defined as

\begin{equation}
\mu_i=\frac{\sigma_{B-L}\times BR(h \to i)_{B-L}}{\sigma_{SM}\times BR(h \to i)_{SM}} ,
\end{equation}

\noindent where $i$ denotes a possible final state of the Higgs boson decay, for example
$b\bar b, W^+W^-, ZZ, gg$ and $\gamma\gamma$.

Fixing the Higgs boson mass to the measured value and considering the decays $h \to \gamma\gamma$,
$h \to ZZ$, $h \to W^+W^-$, $h \to b\bar b$ and $h \to \tau^+\tau^-$, the ATLAS collaboration
report \cite{ATLAS1} a signal strength of

\begin{equation}
\mu=1.18^{+0.15}_{-0.14}.
\end{equation}

\noindent The corresponding CMS collaboration result \cite{CMS1} is

\begin{equation}
\mu=1.00\pm 0.13.
\end{equation}

\noindent Good consistency is found, for both experiments, across different decay modes and analyses categories
related to different production modes.

In the B-L model, the modifications of the $hf\bar f$ (the SM fermions pair) and $hVV$ $(V=W, Z)$ couplings
can give the extra contributions to the Higgs boson production processes. On the other hand, the loop-induced
couplings, such as $h\gamma\gamma$ and $hgg$, could also be affected. Finally, beside the effects already seen
in the Higgs-strahlung channel due to the couplings Eqs. (13), (14) and the functions given by Eq. (17), the
exchange of s-channel heavy neutral gauge boson $Z'$ also affected the production cross section. All effects
can modify the signal strengths in a way that may be detectable at the future ILC/CLIC experiments.

In Fig. 11, we show the dependence of the Higgs signal strengths $\mu_i$ $(i= b\bar b, \gamma\gamma)$ on the
parameter $g'_1$ and $\theta_{B-L}$ for the Higgs-strahlung process $e^{+}e^{-}\rightarrow (Z, Z') \to Zh$,
where (a) and (b) denote the Higgs signal strengths $\mu_{b\bar b}$ and $\mu_{\gamma\gamma}$, respectively.

Using $\theta_{B-L}=10^{−3}$ for the mixing angle and $M_h=125\hspace{1mm}GeV$ for the Higgs boson mass, the
following bound on the signal strength is obtained:

\begin{equation}
\mu=1.2^{+0.12}_{-0.16},
\end{equation}

\noindent which is consistent with that obtained for the ATLAS \cite{ATLAS1} and CMS \cite{CMS1} collaborations,
Eqs. (23) and (24), respectively.

In conclusion, we consider the $Z'$ heavy gauge boson of the B-L model as a Higgs boson factory, through
the Higgs-strahlung process $e^+e^- \to (Z, Z')\to Zh$. We find that the future linear $e^+e^-$ colliders
experiments such as the ILC and CLIC could test the B-L model by measuring the cross section of the process
$e^+e^- \to Zh$, and it would be possible to perform precision measurements of the $Z'$ gauge boson and of the
$h$ Higgs boson, as well as of the parameters of the model $\theta_{B-L}$ and $g'_1$, complementing other
studies on the B-L model and on the Higgs-strahlung process. The SM expression for the cross section of the reaction
$e^+e^- \to Zh$ can be obtained in the decoupling limit, that is to say, when $\theta_{B-L}= 0$ and $g'_1=0$,
in this case the terms that depend on $\theta_{B-L}$ and $g'_1$ in (19) are zero and (19) is reduced to the
expression given in Refs. \cite{Ellis,Barger} for the standard model. We also studied the dependence of the Higgs
signal strengths $(\mu)$ on the parameters $g'_1$ and $\theta_{B-L}$ of the $U(1)_{B-L}$ model for the Higgs-stralung
process $e^+e^- \to Zh$. We obtain a bound on $(\mu)$, which is consistent with that obtained for the ATLAS \cite{ATLAS1}
and CMS \cite{CMS1} collaborations. In addition, the analytical and numerical results for the total cross section have
never been reported in the literature before and could be of relevance for the scientific community.

\vspace{8mm}

\begin{center}
{\bf Acknowledgments}
\end{center}

We acknowledge support from CONACyT, SNI, PROMEP and PIFI (M\'exico).

\newpage

\begin{figure}[t]
\centerline{\scalebox{0.75}{\includegraphics{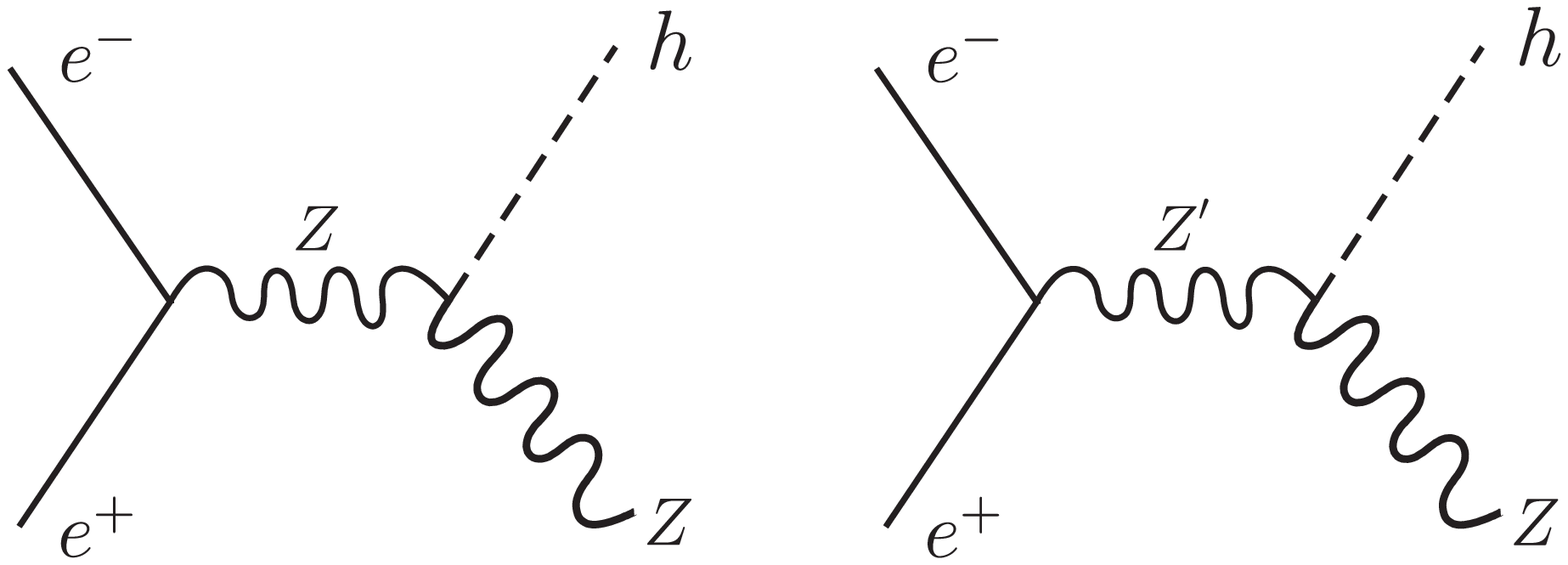}}}
\caption{ \label{fig:con-2Z2gamma} Feynman diagram for the Higgs-strahlung process
$e^+e^-\to Zh$ in the B-L model.}
\end{figure}

\begin{figure}[t]
\centerline{\scalebox{0.72}{\includegraphics{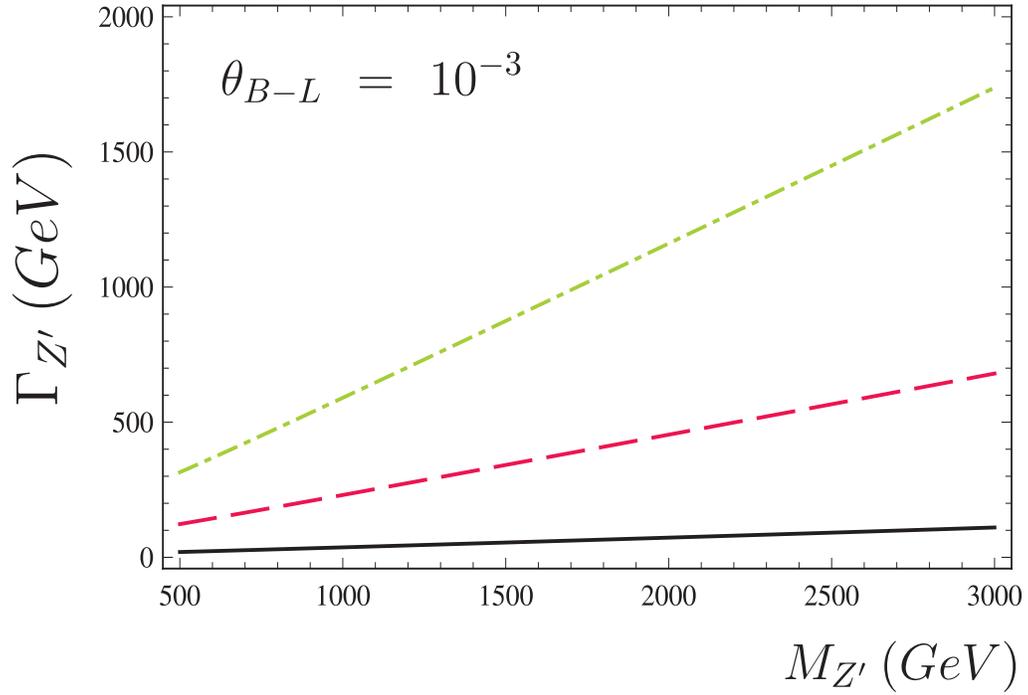}}}
\caption{ \label{fig:con-2Z2gamma} $Z'$ width as a function of $M_{Z'}$ for fixed values
of $g'_1$. Starting from the bottom, the curves are for $g'_1=0.2, 0.5, 0.8$,
respectively.}
\end{figure}

\begin{figure}[t]
\centerline{\scalebox{0.72}{\includegraphics{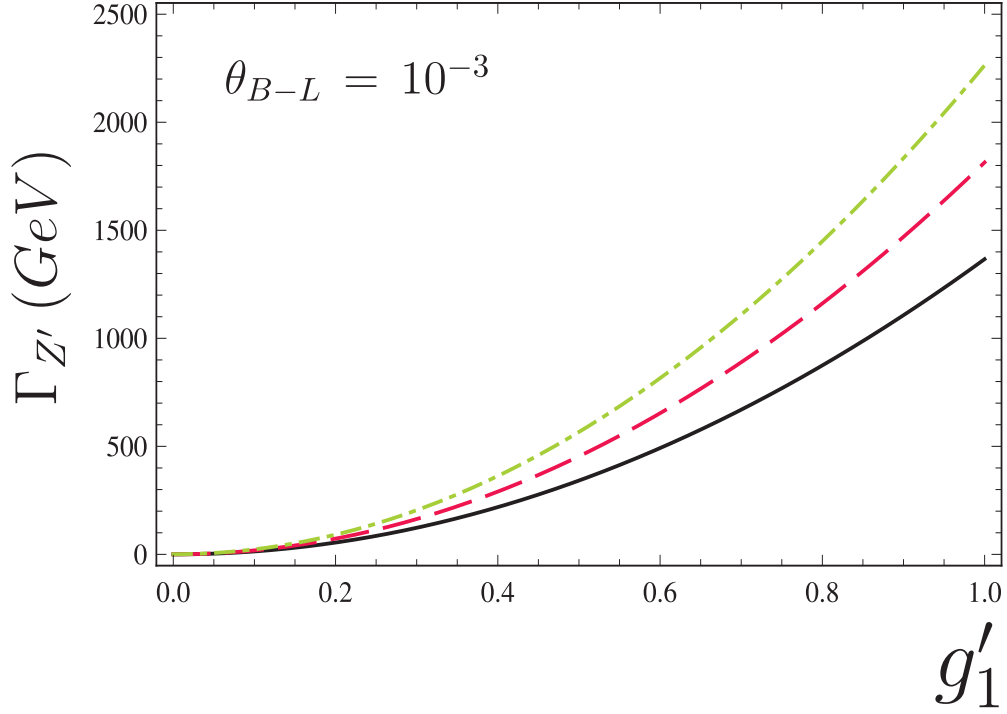}}}
\caption{ \label{fig:con-2Z2gamma} $Z'$ width as a function of $g'_1$ for fixed values
of $M_{Z'}$. Starting from the bottom, the curves are for $M_{Z'}=1500, 2000, 2500$$\hspace{0.8mm}GeV$,
respectively.}
\end{figure}

\begin{figure}[t]
\centerline{\scalebox{0.72}{\includegraphics{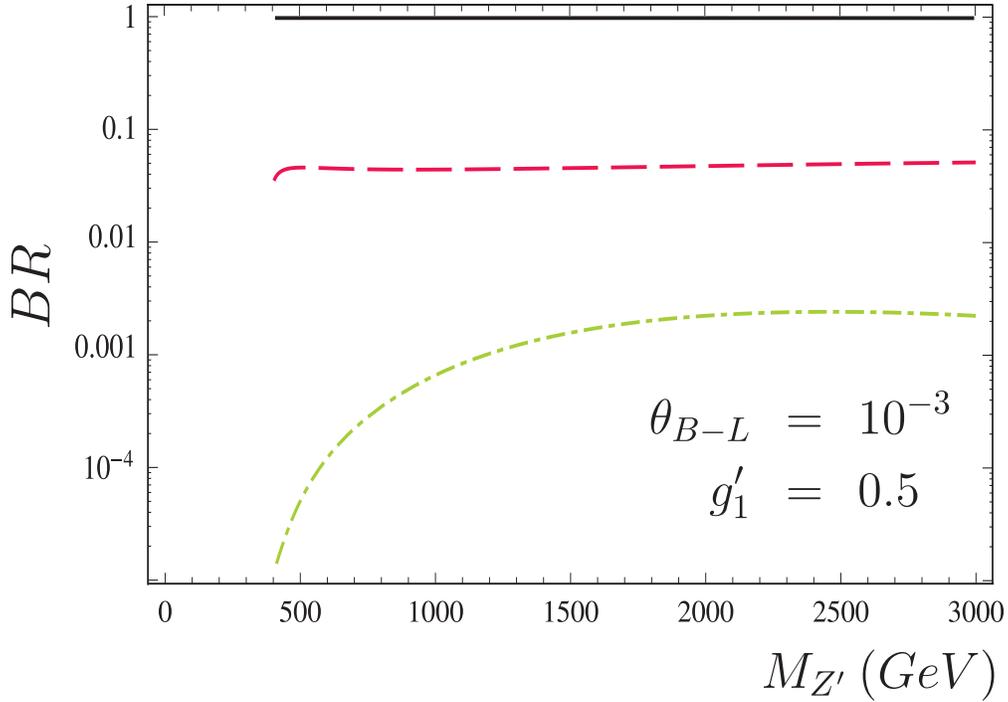}}}
\caption{ \label{fig:con-2Z2gamma} Branching ratios as a function of $M_{Z'}$.
Starting from the top, the curves are for the $BR(f\bar f)$, $BR(Zh)$ and $BR(W^+W^-)$, respectively.}
\end{figure}

\begin{figure}[t]
\centerline{\scalebox{0.65}{\includegraphics{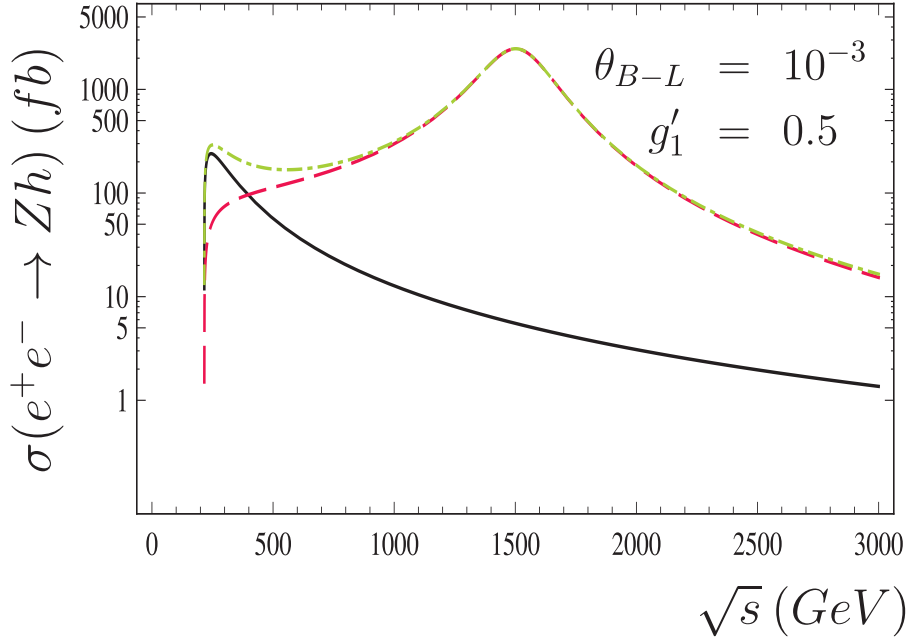}}}
\caption{ \label{fig:con-2Z2gamma} The total cross sections of the production processes
$e^+e^-\to Zh$ as a function of the collision energy for $M_{Z'}=1500$\hspace{0.8mm}$GeV$
and $M_h=125$\hspace{0.8mm}$GeV$. The curves are for the first term of Eq. (19) (solid line),
second term of Eq. (19) (dashed line) and the dot-dashed line correspond to the total cross
section of the process $\sigma(e^+e^− \to Zh)$, respectively.}
\end{figure}

\begin{figure}[t]
\centerline{\scalebox{0.65}{\includegraphics{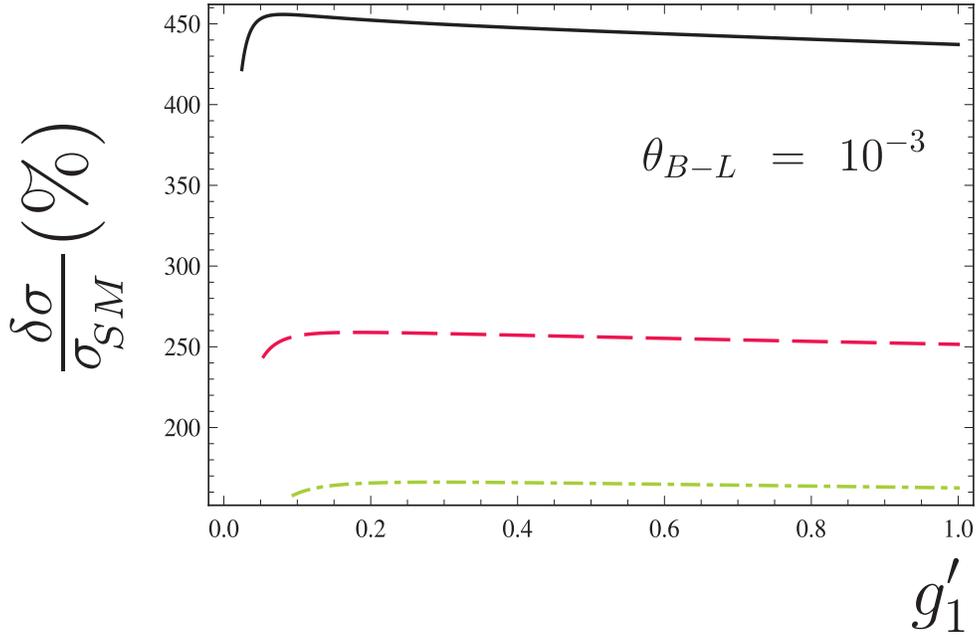}}}
\caption{ \label{fig:con-2Z2gamma} The relative correction $\delta \sigma/\sigma_{SM}$ as a function
of $g'_1$. Starting from the top, the curves are for $M_{Z'}=1500, 2000, 2500\hspace{0.8mm}GeV$ and
$\sqrt{s}=1500, 2000, 2500\hspace{0.8mm}GeV$, respectively.}
\end{figure}

\begin{figure}[t]
\centerline{\scalebox{0.7}{\includegraphics{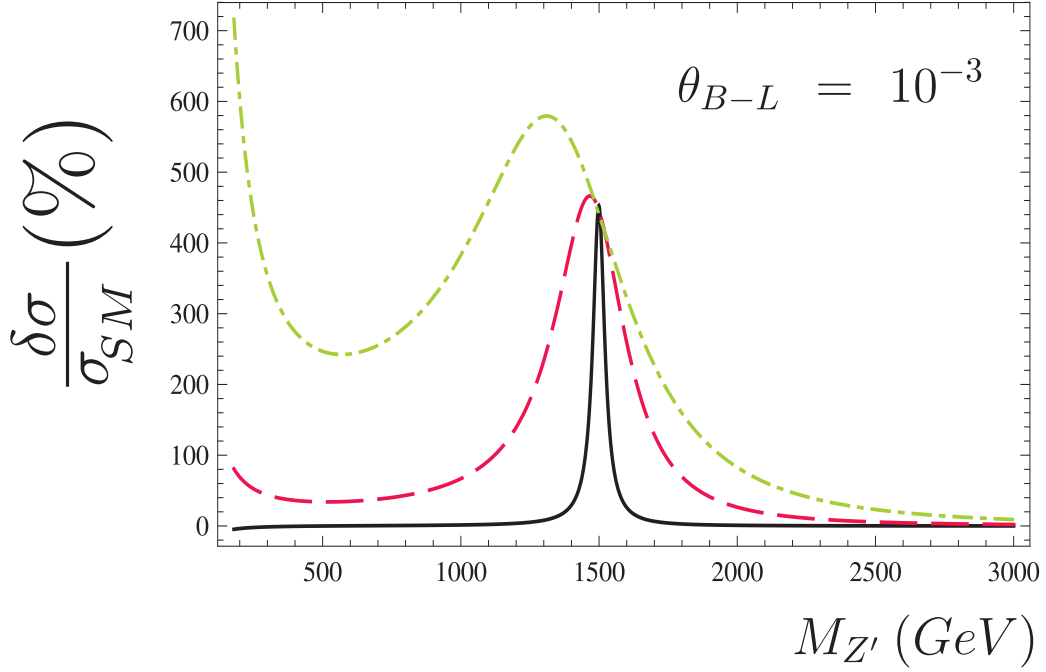}}}
\caption{ \label{fig:con-2Z2gamma} The relative correction $\delta \sigma/\sigma_{SM}$ as a function
of $M_{Z'}$. Starting from the bottom the curves are for $g'_1=0.2, 0.5, 0.8\hspace{0.8mm}GeV$ and
$\sqrt{s}=1500\hspace{0.8mm}GeV$, respectively.}
\end{figure}

\begin{figure}[t]
\centerline{\scalebox{0.7}{\includegraphics{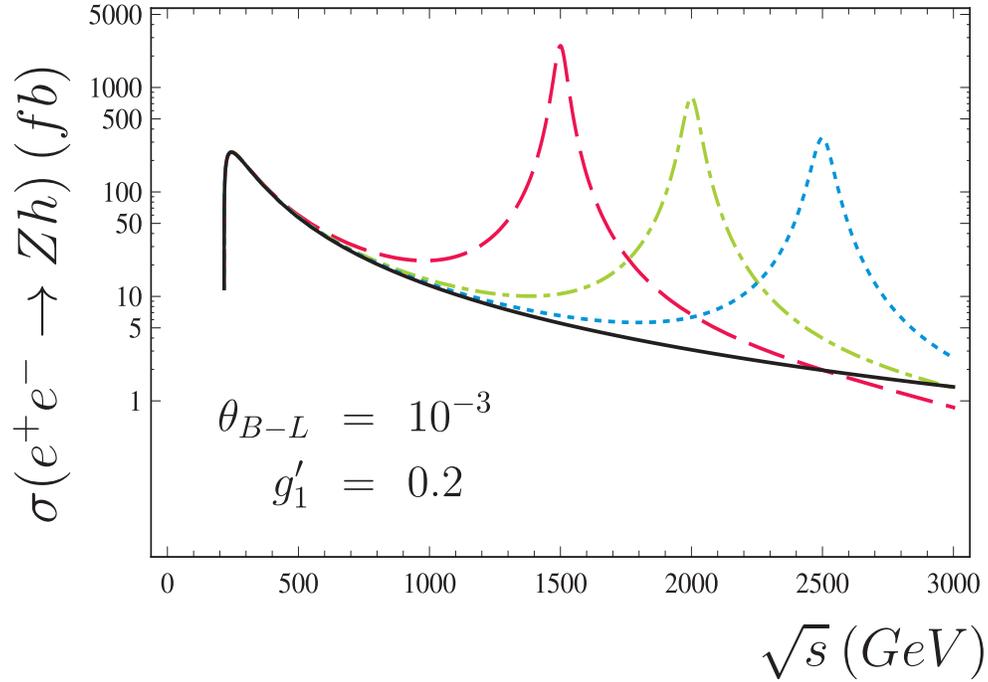}}}
\caption{ \label{fig:con-2Z2gamma} The total cross sections of the production processes $e^+e^-\to Zh$
as a function of the collision energy. The curves are for SM (solid line) and $M_{Z'}=1500, 2000, 2500$\hspace{0.8mm}$GeV$ (dashed-line, dot-dashed line, dotted line). The resonance corresponds to the
$Z'$ new gauge boson.}
\end{figure}

\begin{figure}[t]
\centerline{\scalebox{0.72}{\includegraphics{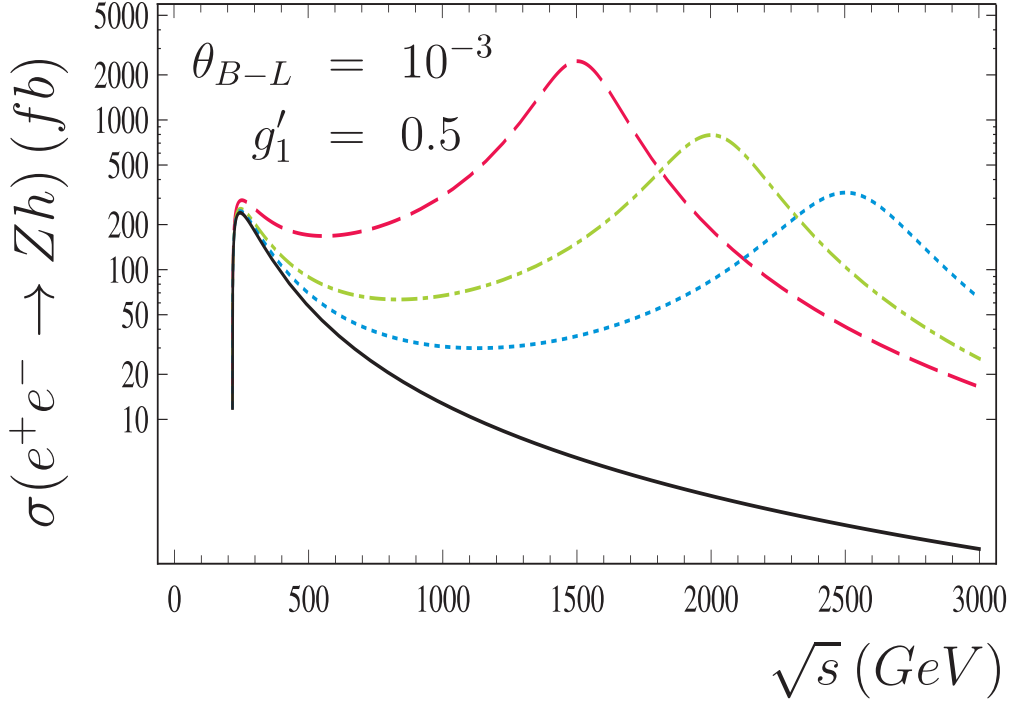}}}
\caption{ \label{fig:con-2Z2gamma} The same as Fig. 8 but for $g'_1=0.5$.}
\end{figure}

\begin{figure}[t]
\centerline{\scalebox{0.7}{\includegraphics{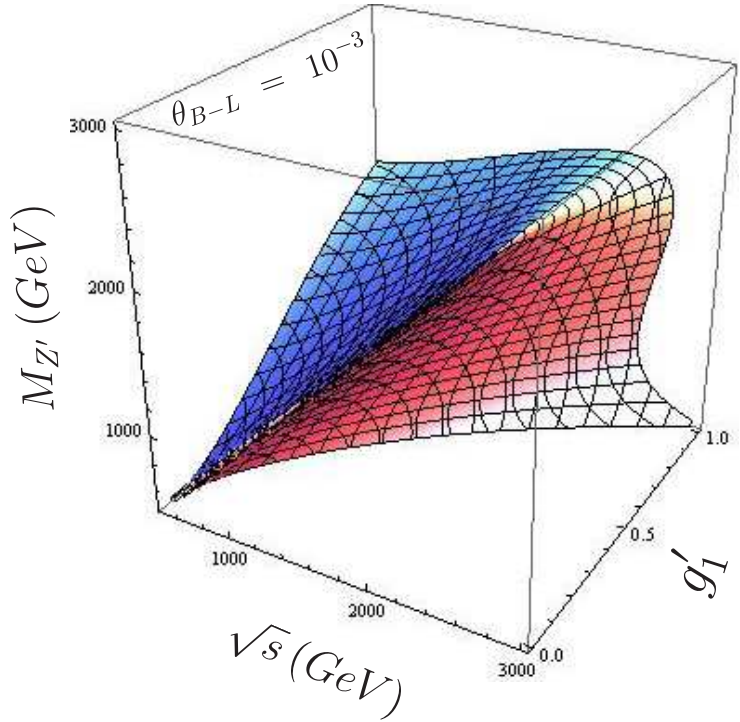}}}
\caption{ \label{fig:con-2Z2gamma} Contour plot 3D for the total cross section $\sigma_{tot}=\sigma_{tot}(\sqrt{s}, M_{Z'}, g'_1)$
of the process $e^+e^- \to Zh$ for $M_h=125$\hspace{0.8mm}$GeV$ and $\theta_{B-L}=10^{-3}$.}
\end{figure}

\begin{figure}[t]
\centerline{\scalebox{0.7}{\includegraphics{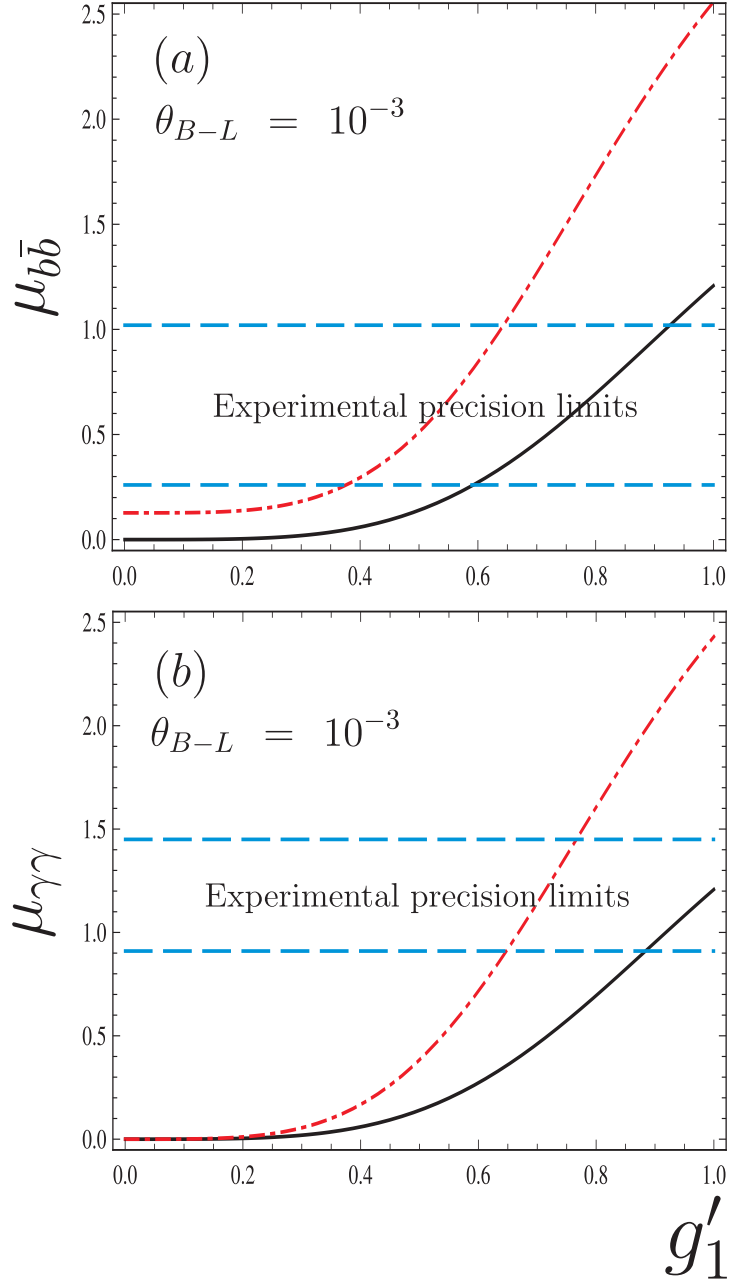}}}
\caption{ \label{fig:con-2Z2gamma} Higgs signal strengths $\mu_i\hspace{2mm}(i=b\bar b, \gamma\gamma)$ for the process
$e^+e^-\to Zh$ as a function of $g'_1$. The dashed lines represent the experimental precision limits and
the solid and dot-dashed lines correspond to the $U(1)_{B-L}$ model with $M_h=125\hspace{1mm}GeV $ and
$\sqrt{s}=500, 1500$\hspace{0.8mm}$GeV$, respectively.}
\end{figure}

\end{document}